\documentclass[aps,preprint,showpacs,preprintnumbers,showkeys,eqsecnum,amsmath,amssymb]{revtex4}

\usepackage{graphics}
\usepackage{graphicx}
\usepackage{txfonts}
\usepackage{dcolumn}
\usepackage{subfigure}
\usepackage{mathrsfs}% Align table columns on decimal point
\usepackage{bm}% bold math
\usepackage{amsmath,amssymb,epsfig,float}
\usepackage{float}

\begin{document}

\title{Holographic Superconductors in Gauss-Bonnet
gravity \\ with Born-Infeld electrodynamics}

\author{ Jiliang {Jing}\footnote{Electronic address:
jljing@hunnu.edu.cn}}
\author{ Liancheng Wang}
\author{ Qiyuan Pan}
\author{ Songbai Chen}
\affiliation{ Institute of Physics and
Department of Physics,
Hunan Normal University, Changsha, Hunan 410081, P. R. China \\
and
\\ Key Laboratory of Low Dimensional Quantum Structures and
Quantum Control of Ministry of Education, Hunan Normal University,
Changsha, Hunan 410081, P. R. China}

\vspace*{0.2cm}
\begin{abstract}

We investigate the holographic superconductors in Gauss-Bonnet gravity with  Born-Infeld electrodynamics. We find that the Gauss-Bonnet constant, the model parameters and the Born-Infeld coupling parameter will affect the formation of the scalar hair, the transition point of the phase transition from the second order to the first order, and
the relation connecting the gap frequency in conductivity with the critical temperature. The combination of the Gauss-Bonnet gravity and the Born-Infeld electrodynamics provides richer physics in  the phase transition and the condensation of the scalar hair.

\end{abstract}



\maketitle

\section{Introduction}

The AdS/CFT correspondence~\cite{Maldacena,polyakov,Witten} relates
a weak coupling gravity theory in an anti-de Sitter space to a
strong coupling conformal field theory in one less dimensions.
Recently it has been applied to condensed matter physics and in
particular to superconductivity \cite{Gubser:2005ih,GubserPRD78}. In the pioneering papers Gubser \cite{Gubser:2005ih,GubserPRD78}
suggested that near the horizon of a charged black hole there is in
operation a geometrical mechanism parameterized by a charged scalar
field of breaking a local $U(1)$ gauge symmetry. Then, the
gravitational dual of the transition from normal to superconducting
states in the boundary theory was constructed. This dual consists of a system with a black hole and a charged scalar field, in which the
black hole admits scalar hair at temperature lower than a critical
temperature, but does not possess scalar hair at higher
temperatures~\cite{HartnollPRL101}. In this system a scalar
condensate can take place through the coupling of the scalar field
with the Maxwell field. Much attention has been
focused on the application of AdS/CFT correspondence to condensed
matter physics since then
\cite{HartnollJHEP12,HorowitzPRD78,Nakano-Wen,Amado,
Koutsoumbas,Maeda79,Sonner,HartnollRev,HerzogRev,
Ammon:2008fc,Gubser:2009qm,CJ0}.

Recently, it is of great interest to generalize the investigation to
the Einstein-Gauss-Bonnet gravity, which is motivated by the
application of the Mermin-Wagner theorem to the holographic
superconductors. It was found
\cite{Gregory,Pan-Wang,Ge-Wang,Brihaye,Gregory1009} that the higher
curvature corrections in general make the condensation of the scalar field harder to form and give larger corrections to the so-called Horowitz's relation $\omega_{g}/T_c\approx8$ for the conductivity. And then the general
holographic superconductor models in Einstein-Gauss-Bonnet gravity
are constructed and it is observed that different values of
Gauss-Bonnet correction term and model parameters can determine the
order of phase transitions and critical exponents of second-order
phase transitions \cite{Pan-Wang1}. Very recently, the holographic p-wave superconductor models in the Gauss-Bonnet gravity was introduced in the probe limit and the Gauss-Bonnet correction term can also effect the condensation of the vector field \cite{Cai-pGB}.

On the other hand, non-linear electrodynamics has been a subject of research for many
years.  Heisenberg and Euler \cite{euler} noted that quantum electrodynamics predicts that the electromagnetic field behaves non-linearly through the presence of virtual charged particles.
Born and Infeld \cite{born} presented a new classical non-linear theory of  electromagnetism which contains many symmetries common to the Maxwell theory despite its non-linearity. It was found that  the Born-Infeld electrodynamics is the only possible non-linear version of electrodynamics that is invariant under electromagnetic duality transformations \cite{gibb}. Thus, the interest of study for Born-Infeld electrodynamics has been arisen in \cite{Olivera1,Hoffman-Gibbons-Rasheed, Oliveira}.
The static spherically symmetric black holes for the Born-Infeld
electrodynamics coupled to Einstein gravity was derived in Refs. \cite{Hoffman-Gibbons-Rasheed,Oliveira}. Within the framework of AdS/CFT correspondence, we studied the effects of the Born-Infeld electrodynamics on the holographic superconductors in the background of a Schwarzschild-AdS black hole spacetime \cite{Jing-Chen}.

Motivated by the recent studies mentioned above and the fact that,
within the framework of AdS/CFT correspondence, higher-derivative corrections to either gravitational or electromagnetic action in AdS space are expected to modify the dynamics of the strongly coupled dual theory, in this paper we
will investigate the behavior of the holographic superconductors in the Gauss-Bonnet gravity with the Born-Infeld electrodynamics in a five dimensional planar black-hole background, and to see how the combination of the Gauss-Bonnet gravity and the Born-Infeld electrodynamics affect the formation of the scalar hair, the phase transition and Horowitz's relation.

The paper is organized as follows. In Sec. II,  we explore
the scalar condensation in the background of the Gauss-Bonnet black hole by introducing a complex charged scalar field coupling with an electric field obeyed to Born-Infeld electrodynamics. In Sec. III, we study the electrical conductivity and find the ratio of the
gap frequency in conductivity to the critical temperature. We
summarize and discuss our conclusions in the last section.

\section{Scalar condensation}

The Einstein-Gauss-Bonnet theory is the most general Lovelock theory in five and six dimensions and the action is described by
\begin{equation}
I_{grav}=\frac{1}{16\pi G}\int\limits_{\mathcal{M}}d^{d}x\,\sqrt{-g}\,\left[
R-2\Lambda +\hat{\alpha} \,\left( R^{2}-4R_{\mu \nu }R^{\mu \nu }+R_{\mu \nu
\lambda \sigma }R^{\mu \nu \lambda \sigma }\right) \right] \,,
\end{equation}%
where  $\Lambda=-(d-1)(d-2)/(2L^2) $ is the cosmological constant,
$G$ is the gravitational constant, and $\hat{\alpha} $ is the
Gauss-Bonnet coupling constant. The static spacetime of a neutral
black hole in $d$ dimensional Einstein-Gauss-Bonnet gravity is
\cite{Boulware-Deser,Cai-2002,Charmousis:2002rc}
\begin{eqnarray}\label{BH metric}
ds^2=-f(r)dt^{2}+\frac{dr^2}{f(r)}+r^{2}dx_{i}dx^{i},
\end{eqnarray}
with
\begin{eqnarray}
f(r)=\frac{r^2}{2\alpha}\left[1-\sqrt{1-\frac{4\alpha}{L^{2}}
\left(1-\frac{ML^{2}}{r^{d-1}}\right)}~\right],
\end{eqnarray}
where $\alpha=\hat{\alpha}(d-3)(d-4)$ and the constant $M$ is relate
to the black hole horizon by $r_{+}=(ML^{2})^{1/(d-1)}$. In the
asymptotic region ($r\rightarrow\infty$), we have $
f(r)\sim\frac{r^2}{2\alpha}\left(1-\sqrt{1-4\alpha/L^2} \right). $
Thus, we can define the effective asymptotic AdS scale by $ L^2_{\rm
eff}=(2\alpha)/(1-\sqrt{1-4\alpha/L^2}). $ The Hawking temperature
of the black hole is
\begin{eqnarray}
\label{Hawking temperature} T=\frac{(d-1)r_{+}}{4\pi L^{2}}\ .
\end{eqnarray}

We now consider the Born-Infeld electrodynamics and the charged scalar field coupled via a generalized Lagrangian
\begin{eqnarray}\label{System}
S=\int d^{d}x\sqrt{-g}\left[
\frac{1}{b^{2}}\left( 1
-\sqrt{1+\frac{b^{2} F^{2}}{2}}\right)
-\frac{1}{2}\partial_{\mu}\tilde{\psi}\partial^{\mu}\tilde{\psi}
-\frac{1}{2}m^2\tilde{\psi}^2-\frac{1}{2}|\mathfrak{F}(\tilde{\psi})|(\partial_{\mu}p-A_{\mu})
(\partial^{\mu}p-A^{\mu}) \right] \ ,\nonumber\\
\end{eqnarray}
where $\mathfrak{F}$ is taken as $
\mathfrak{F}(\tilde{\psi})=\tilde{\psi}^{2}
+c_{\gamma}\tilde{\psi}^{\gamma}+c_{4}\tilde{\psi}^{4} $ with the
model parameters $c_{\gamma}$, $\gamma$ and $c_{4}$ in order to
introduce a general class of gravity duals to superconducting
theories that exhibit both first and second-order phase transitions
at finite temperature in strongly interacting systems
\cite{FrancoPRD,Franco}, and it reduces to the model considered in
\cite{Jing-Chen} if $c_{\gamma}$ and $c_{4}$ are zero. It should be
noted that $b$ is the Born-Bonnet coupling parameter and the
Born-Infeld electrodynamics will reduce to the Maxwell case in the
weak-coupling limit $b\rightarrow 0$. We can use the gauge freedom
to fix $p=0$ and take $\psi\equiv\tilde{\psi}$, $A_{t}=\phi$ where
$\psi$, $\phi$ are both real functions of $r$ only. Then the
equations of motion are given by
\begin{eqnarray}\label{Psi}
&&\psi^{\prime\prime}+\left(
\frac{f^\prime}{f}+\frac{d-2}{r}\right)\psi^\prime
+\frac{\phi^2}{f^2}\left(\psi+\frac{\gamma}{2}c_\gamma \psi^\gamma+2c_4\psi^3 \right)-\frac{m^2}{f}\psi=0\,,
\\ \label{Phi} &&
\left(\phi^{\prime\prime}+\frac{d-2}{r}\phi^\prime\right)
\left(1-b^2 \phi^{\prime 2}\right)+b^2\phi^{\prime 2}\phi^{\prime \prime}-\left(1-b^2 \phi^{\prime 2}\right)^{\frac{3}{2}}\left(\psi^2+c_\gamma \psi^\gamma+c_4\psi^4 \right)\frac{\phi}{f}=0~,
\end{eqnarray}
where  a prime denotes the derivative with respect to $r$. At the
event horizon $r=r_+$, we must have
\begin{eqnarray}
 \psi(r_{+})&=&-\frac{(d-1) }{ m^{2} L^2}\psi^\prime(r_{+}),~~~~  \phi(r_{+})=0,
\end{eqnarray} and at the asymptotic AdS region
($r\rightarrow\infty$), the solutions behave like
\begin{eqnarray}
\psi=\frac{\psi_{-}}{r^{\lambda_{-}}}+\frac{\psi_{+}}{r^{\lambda_{+}}}\,,\hspace{0.5cm}
\phi=\mu-\frac{\rho}{r^{d-3}}\,, \label{infinity}
\end{eqnarray}
with
\begin{eqnarray}
\lambda_\pm=\frac{1}{2}\left[(d-1)\pm\sqrt{(d-1)^{2}+4m^{2}L_{\rm
eff}^2}~\right]\,, \label{LambdaZF}
\end{eqnarray}
where $\mu$ and $\rho$ are interpreted as the chemical potential and
charge density in the dual field theory respectively. We take
$\psi_{-}=0$ because we can impose boundary condition that either
$\psi_{+}$ or $\psi_{-}$ vanishes
\cite{HartnollPRL101,HartnollJHEP12}, and we will focus on  $d=5$
and $m^2L^2=-3$ here.  Thus, the scalar condensate is now described
by the operator $\langle{\cal O}_{+}\rangle=\psi_{+}$. In what
following we will present a detail analysis of the condensation of
the operator $\langle{\cal O}_{+}\rangle$ by taking numerical
integration of  the equations (\ref{Psi}) and (\ref{Phi})    from
the horizon out to the infinity with the boundary  conditions
mentioned above.
\begin{figure}[H]
\includegraphics[scale=0.72]{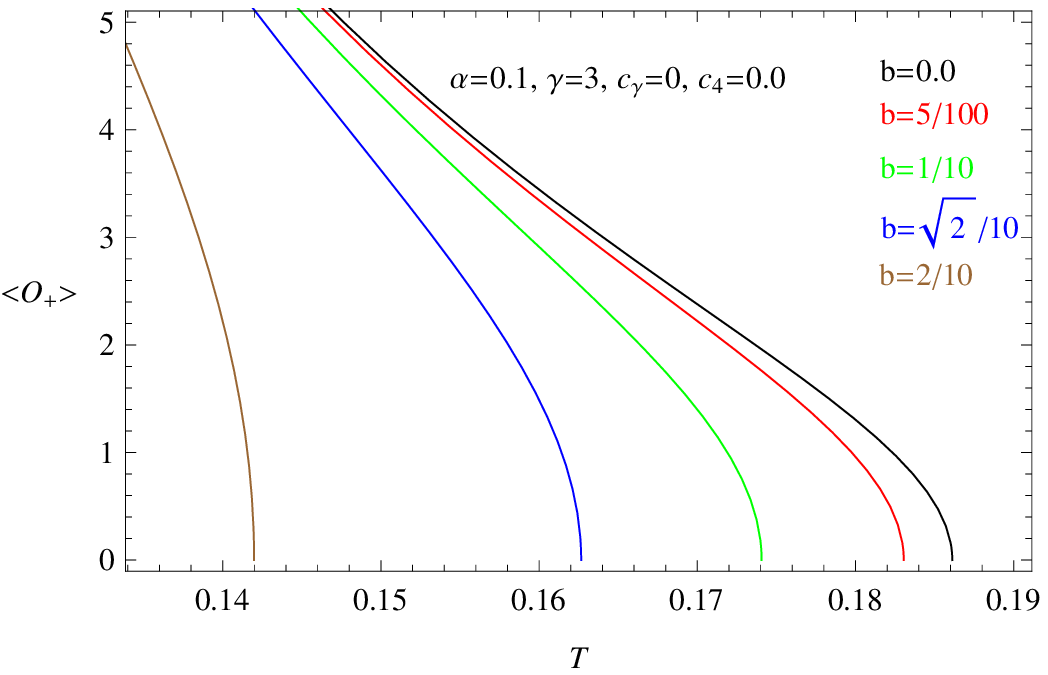}\hspace{0.2cm}%
\includegraphics[scale=0.72]{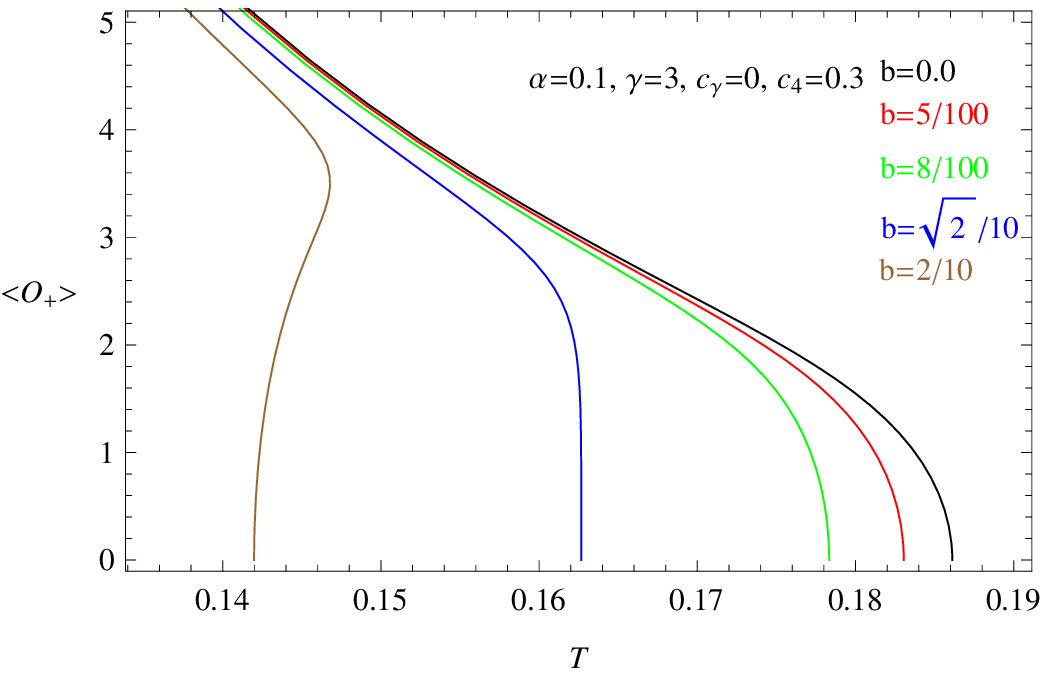}\\ \vspace{0.2cm}%
\includegraphics[scale=0.72]{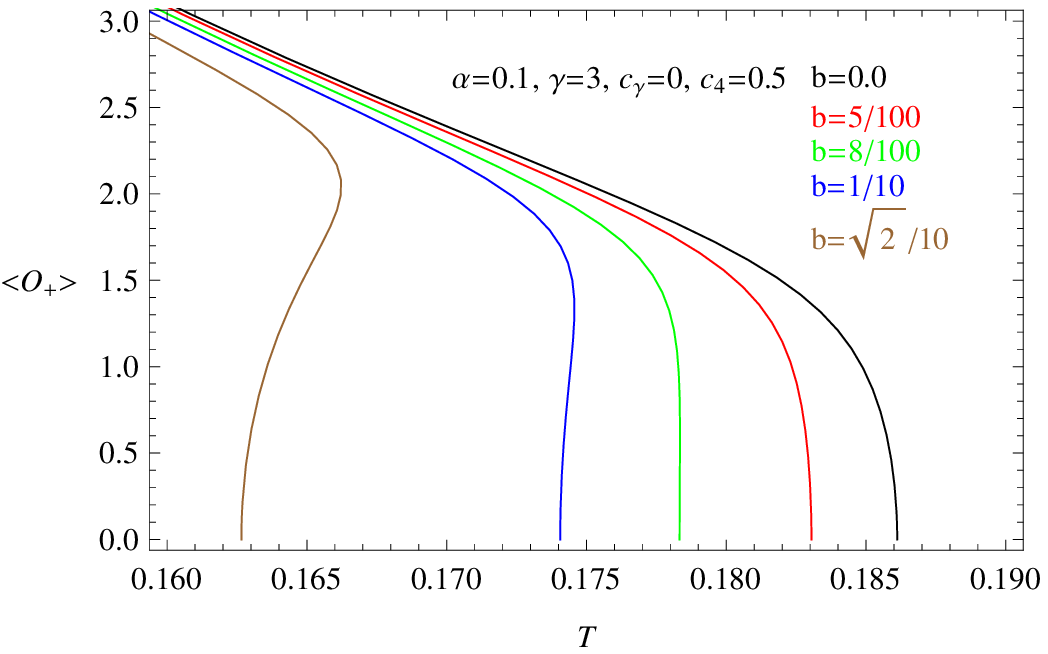}\hspace{0.2cm}
\includegraphics[scale=0.72]{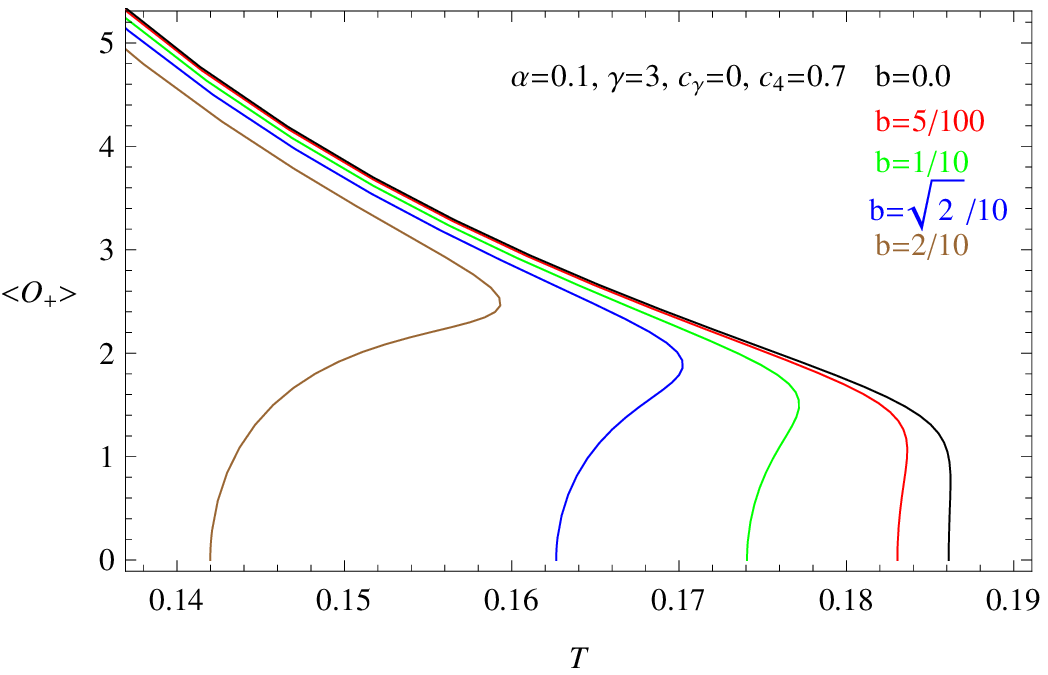}\\ \vspace{0.0cm}%
\includegraphics[scale=0.72]{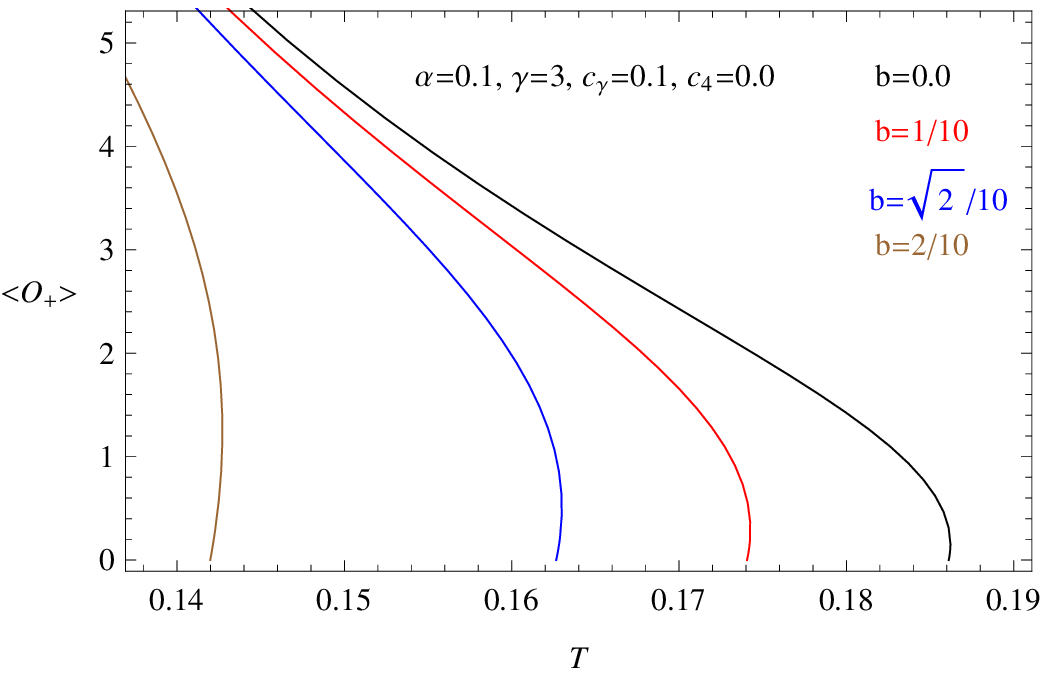}\hspace{0.2cm}%
\includegraphics[scale=0.72]{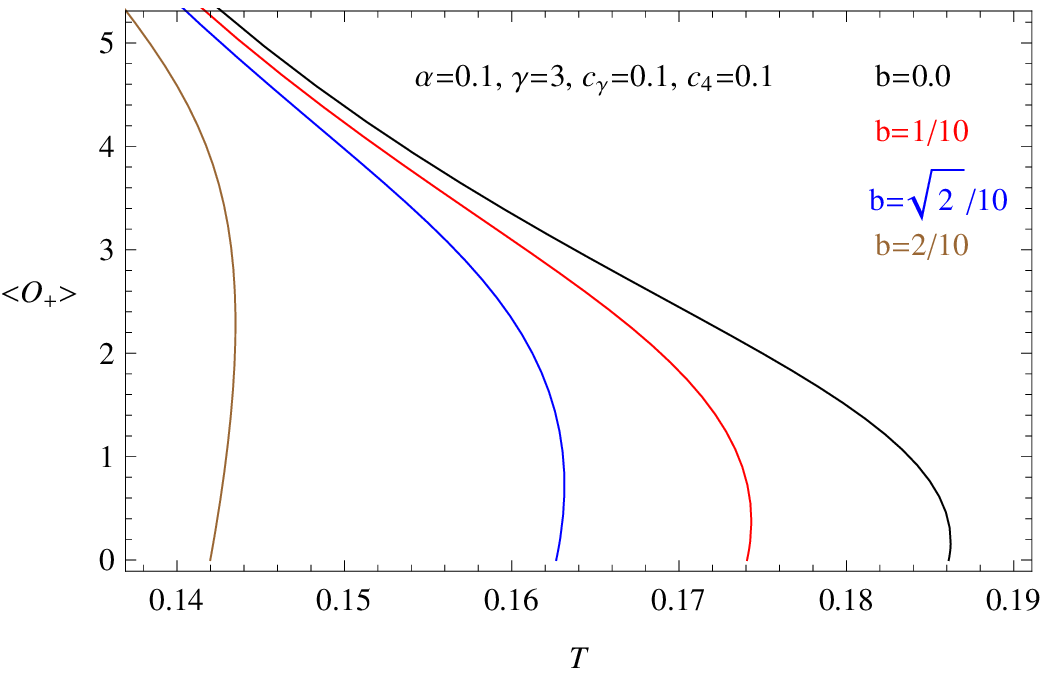}\\ \vspace{0.2cm}%
\includegraphics[scale=0.72]{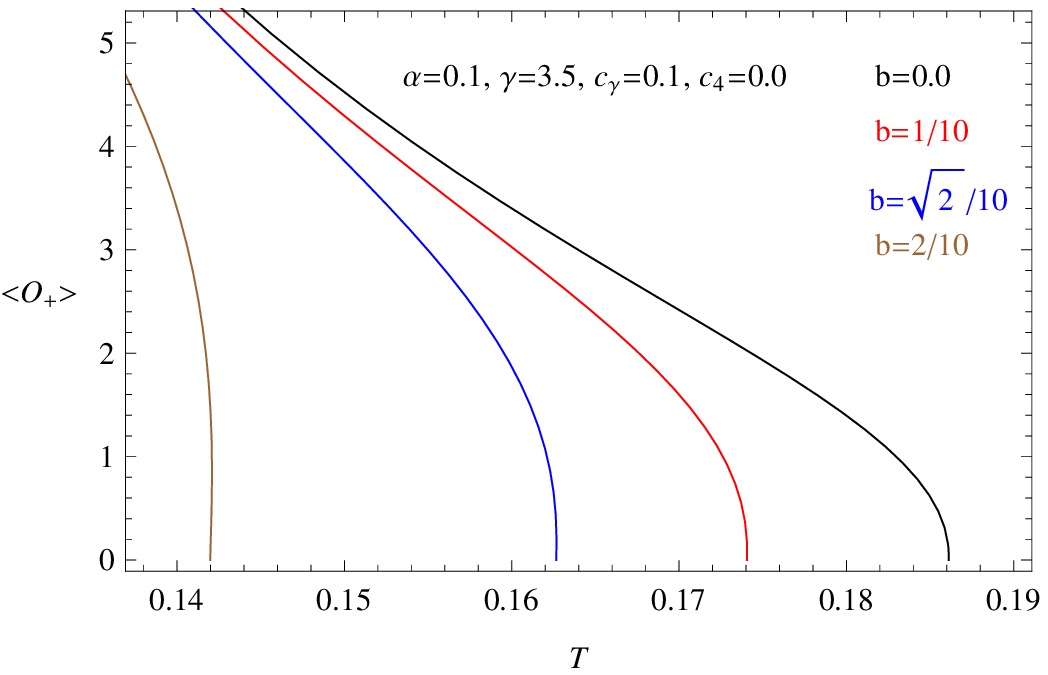}\hspace{0.2cm}
\includegraphics[scale=0.72]{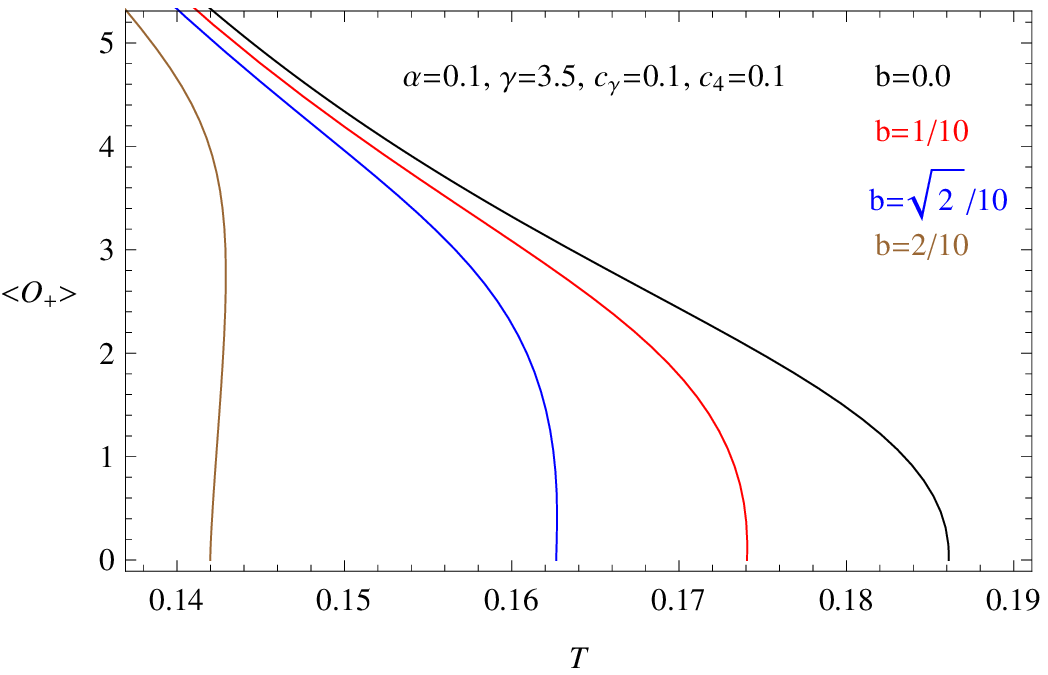}
\caption{\label{condensate} (Color online) The condensate
$\langle{\cal O}_{+}\rangle$  as a function of temperature with
fixed value $\alpha=0.1$ for different values of the model
parameters ($c_\gamma,~\gamma, c_4$) and Born-Infeld coupling
parameter $b$, which shows that a different values of these
parameters not only change the formation of the scalar hair, but
also separate the first- and second-order phase transition. }
\end{figure}

\begin{table}[ht]
\begin{center}
\caption{\label{Tc-abc} The critical values of $T_c$ and $b_c$ for
different $\alpha$ and $c_{4}$, which can separate the first- and
second-order phase transitions for the simple model $\mathfrak{F}
(\psi) = \psi^2+c_4\psi^4$.  The word ``No" in the table
corresponds to the inexistence of the critical point.}
\begin{tabular}{c | c | c | c | c |c}
         \hline
$ $ &$c_{4}=0$ & $c_{4}=0.3$~~&$c_{4}=0.5$~~& $c_{4}=0.7$~~& $c_{4}=1.0$
        \\ \hline
$\alpha=0.1~~$& No &$b_c=\frac{\sqrt{2}}{10}$~~$T_c=0.16253$&
 $b_c=\frac{1}{10}$~~$T_c=0.17456$&$b_c=0$~~~~~$T_c=0.18622$& No
          \\
$\alpha=0.0~~$& No &$b_c=\frac{2}{10}$~~~$T_c=0.16566$&
 $b_c=\frac{\sqrt{2}}{10}$~~$T_c=0.18118$&$b_c=\frac{1}{10}$
 ~~$T_c=0.18973$& No
 \\
 $\alpha=-0.1$& No &$b_c=\frac{\sqrt{7}}{10}$~~$T_c=0.16824$&
 $b_c=\frac{2}{10}$~~$T_c=0.18454$&$b_c=\frac{\sqrt{2}}{10}$
 ~~$T_c=0.19636$&$b_c=0$~~$T_c=0.20636$
          \\
        \hline
\end{tabular}
\end{center}
\end{table}

We present in Fig.
\ref{condensate} the influence of the parameters  ($c_\gamma,~\gamma,~c_4$) and $b$ on the condensation with fixed values $ m^2 L^2=-3$ and $\alpha=0.1$. In fact, the different choices of $\alpha$ can not qualitatively change our results. We know from the figure that the Born-Infeld coupling parameter and the model parameters have obvious different effects on the critical temperature. If we fix the model parameters ($c_\gamma,~\gamma,~c_4$), we note that the critical temperature becomes smaller as the Born-Infeld coupling parameter $b$ increases for two types of phase transitions, i.e., the scalar hair can be formed harder for the larger $b$.
However, the story is completely different if we fix the Born-Infeld coupling parameter $b$. For the cases of second phase transition, the critical temperature keeps as a constant with the increase of ($c_\gamma,~\gamma,~c_4$). That is to say, the formation of the scalar hair does not affect by model parameter ($c_\gamma,~\gamma,~c_4$). But for the cases of first phase transition, the critical temperature is larger with the increase of ($c_\gamma,~c_4$) or decrease of $\gamma$,  which means that the scalar hair can be formed easier for the larger model parameter ($c_\gamma,~c_4$) or smaller $\gamma$. It should be pointed out that we define the critical temperature $T_{c}$ just as Franco {\it et al} for the first phase transition \cite{Franco}.

From Fig. \ref{condensate} we also find that, for a simple model
$\mathfrak{F}(\psi)=\psi^{2}+c_{4}\psi^{4}$ with fixed $c_4$
($c_4>0)$, there is a phase transition from the second order to the
first one as we increase value of $b$.  In table \ref{Tc-abc}, we
list the critical values of $b_{c}$ and  $T_{c}$ which separate the
second order and the first order phase transitions for selected
$\alpha$ and $c_4$.   Note that the word ``No" in this table
corresponds to the inexistence of the critical values of $b_{c}$ and
$T_{c}$, i.e., the phase transition is always of the second order if
$c_{4}=0$ with different $\alpha$ but the first order if $c_{4}=1.0$
with $\alpha=0.0$ and $0.1$. We learn from the figure and the table
that both $b_c$ and $T_c$ decrease as $\alpha$ increases for fixed
$c_{4}$, and $b_c$ decreases but $T_c$ increases as $c_{4}$
increases for fixed $\alpha$. Thus, the Gauss-Bonnet constant
$\alpha$, the model parameter $c_4$  and the Born-Infeld coupling
parameter $b$  provide richer physics in the phase transition.

\section{Electrical Conductivity }

In the study of (2+1) and (3+1)-dimensional superconductors,
Horowitz {\it et al.} \cite{HorowitzPRD78} got a universal relation
connecting the gap frequency in conductivity with the critical
temperature, which is described by
\begin{eqnarray}
\frac{\omega_g}{T_c}\approx 8,
\end{eqnarray}
with deviations of less than $8\%$. This is roughly twice the BCS
value 3.5 indicating that the holographic superconductors are
strongly coupled.  However, the authors in Refs. \cite{Pan-Wang,Gregory}
found that this relation is not stable in the presence of the
Gauss-Bonnet correction terms. We now examine this relation for the
Gauss-Bonnet gravity with the Born-Infeld electrodynamics.

In order to compute the electrical conductivity, we should study the Born-Infeld electromagnetic perturbation in this Gauss-Bonnet black hole background, and then calculate the linear response to the
perturbation. In the probe approximation, the effect of the
perturbation of metric can be ignored. Assuming that the
perturbation of the vector potential is translational symmetric and
has a time dependence as $\delta A_x=A_x(r)e^{-i\omega  t}$, we find that the motion equation for the Born-Infeld electrodynamics in the Gauss-Bonnet black hole background reads
\begin{eqnarray}
&&\left(A_{x}^{\prime\prime}+\frac{f^\prime}{f}A_{x}^\prime
+\frac{d-4}{r}A_{x}^{\prime}+\frac{\omega^2}{f^2}A_{x}\right)\left(1-b^2 \phi^{\prime 2}\right)\nonumber \\
&&+b^2\phi^{\prime }\phi^{\prime \prime}A_{x}^{\prime}
-\left(1-b^2 \phi^{\prime 2}\right)^{\frac{3}{2}}\left(\psi^2+c_\gamma \psi^\gamma+c_4 \psi^4\right)\frac{A_{x}}{f}=0 \; .
\label{Maxwell Equation}
\end{eqnarray}
From Eq. (\ref{Maxwell Equation}), an
ingoing wave  boundary condition near the horizon is given by
\begin{eqnarray}
A_{x}(r)\sim f(r)^{-\frac{i \omega L^2}{(d-1) r_+}}, \end{eqnarray} and in the asymptotic AdS region ($r\rightarrow\infty$),  the general behavior for $d=5$ should be \cite{Gregory1009}
\begin{eqnarray}\label{Maxwell boundary}
A_{x}=L_e^{-1/2} A^{(0)}+\frac{L_e^{3/2}}{r^2}\left(A^{(2)}-\frac{1}{2} \ln \frac{r}{L} \partial_t^2 A^{(0)}  \right).
\end{eqnarray}
Then the conductivity can be expressed as \cite{Gregory1009}
\begin{eqnarray}\label{GBConductivity}
\sigma=\frac{2A^{(2)}}{i\omega A^{(0)}}+\frac{i\omega}{2} -i \omega \log\frac{L_e}{L} \ ,
\end{eqnarray}
where the factor of $L_e^{-1/2}$ ensures that the gauge fields $A^{(n)}_\mu$ have the correct dimensionality. We can obtain the
conductivity by solving the motion equation (\ref{Maxwell Equation}) numerically for the general forms of function $\mathfrak{F}(\psi)=\psi^{2} +c_{\gamma}\psi^{\gamma}+c_4\psi^4$. Here we also focus our attention on the case for $m^{2}L^{2}=-3$.

\begin{table}[ht]
\caption{\label{ConductivityTc} The ratio $\omega_{g}/T_{c}$ for
different values of the Gauss-Bonnet constant $\alpha$, the model parameter $c_\gamma$ and the Born-Infeld coupling parameter $b$ with $m^{2}L^2=-3$ and $\gamma=3$.}
\begin{tabular}{|c|c|c|c||c|c|c|}   \hline
 \multicolumn{4}{|c||}{$c_{\gamma}=0$} & \multicolumn{3}{c|} {$c_{\gamma}=0.1$} \\
         \hline
~ &~~b=0.0~~&~~b=0.1~~&~~b=0.2~~ &~~b=0.0~~&~~b=0.1~~&~~b=0.2~~
          \\
        \hline
~~$\alpha=0.1~~$~~ & ~~$8.5$~~ & ~~$9.0$~~ & ~~$10.0$~~
 & ~~$  9.3 $~~ & ~~$  9.7  $~~ & ~~$ 11.6  $~~
          \\
        \hline
$\alpha=0.0~~$ & $ 7.7 $ & $8.1 $ & $8.7 $
 & ~~$  8.4 $~~ & ~~$  8.7 $~~ & ~~$ 9.7  $~~
          \\
        \hline
$\alpha=-0.1$ & $7.3$ & $7.6$ & $7.9$
 & ~~$ 7.8  $~~ & ~~$  8.1 $~~ & ~~$  9.0 $~~
          \\
         \hline
\end{tabular}
\end{table}

\begin{figure}[H]
\includegraphics[scale=0.5]{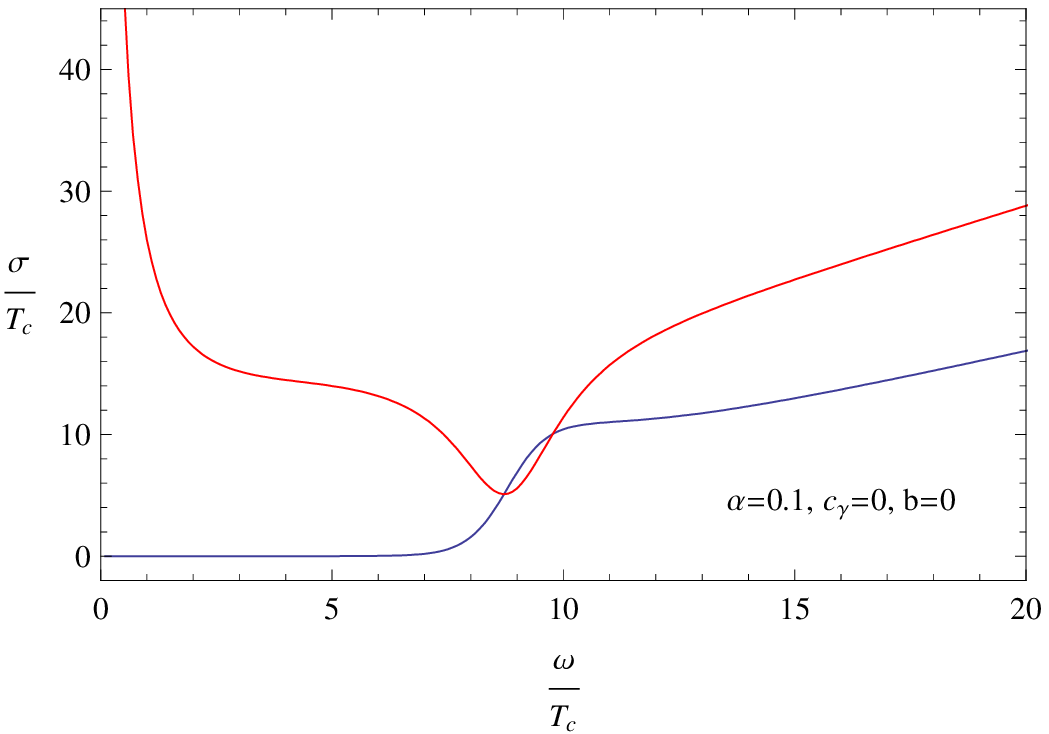}\hspace{0.2cm}%
\includegraphics[scale=0.5]{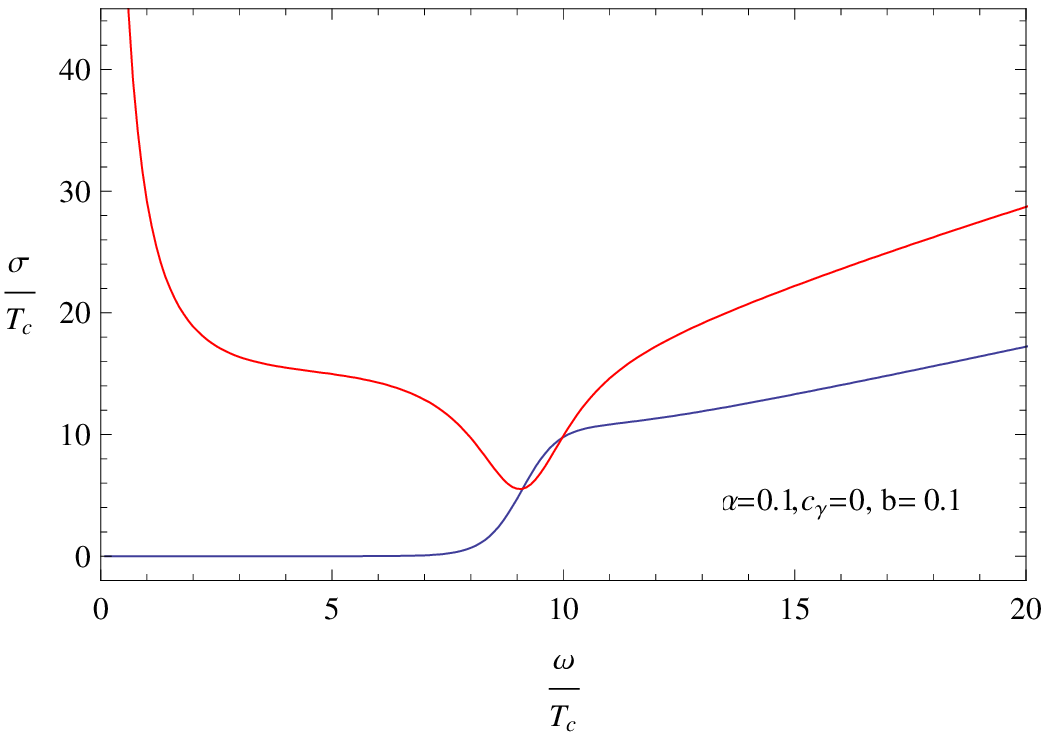}\hspace{0.2cm}%
\includegraphics[scale=0.5]{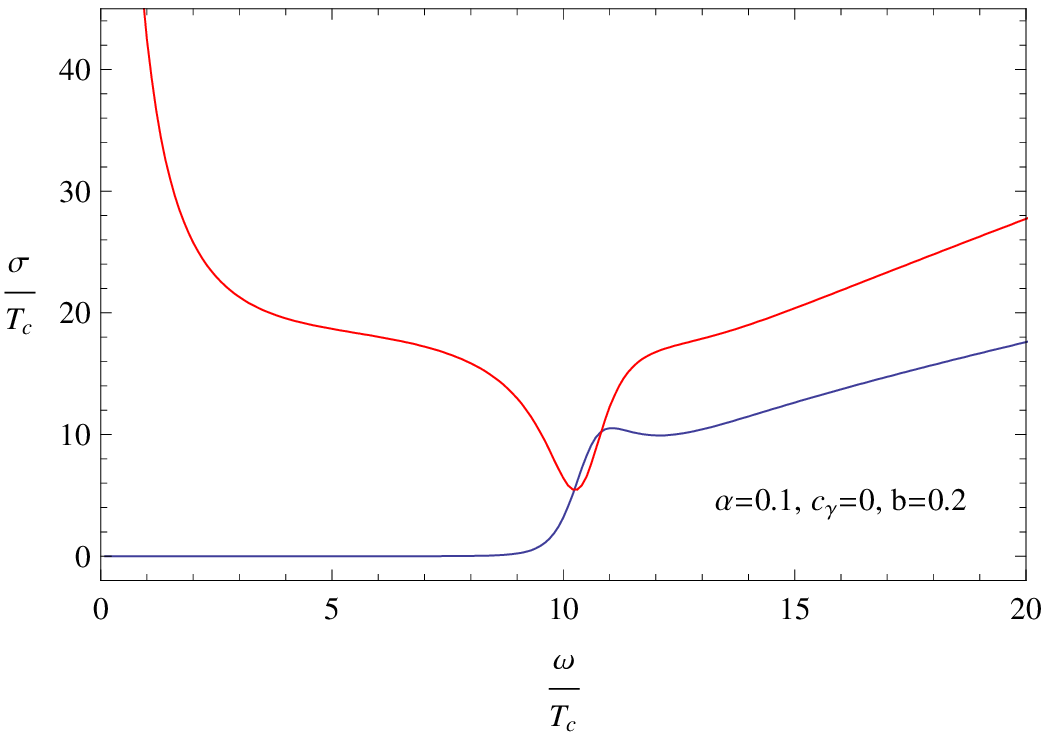}\\ \vspace{0.0cm}
\includegraphics[scale=0.5]{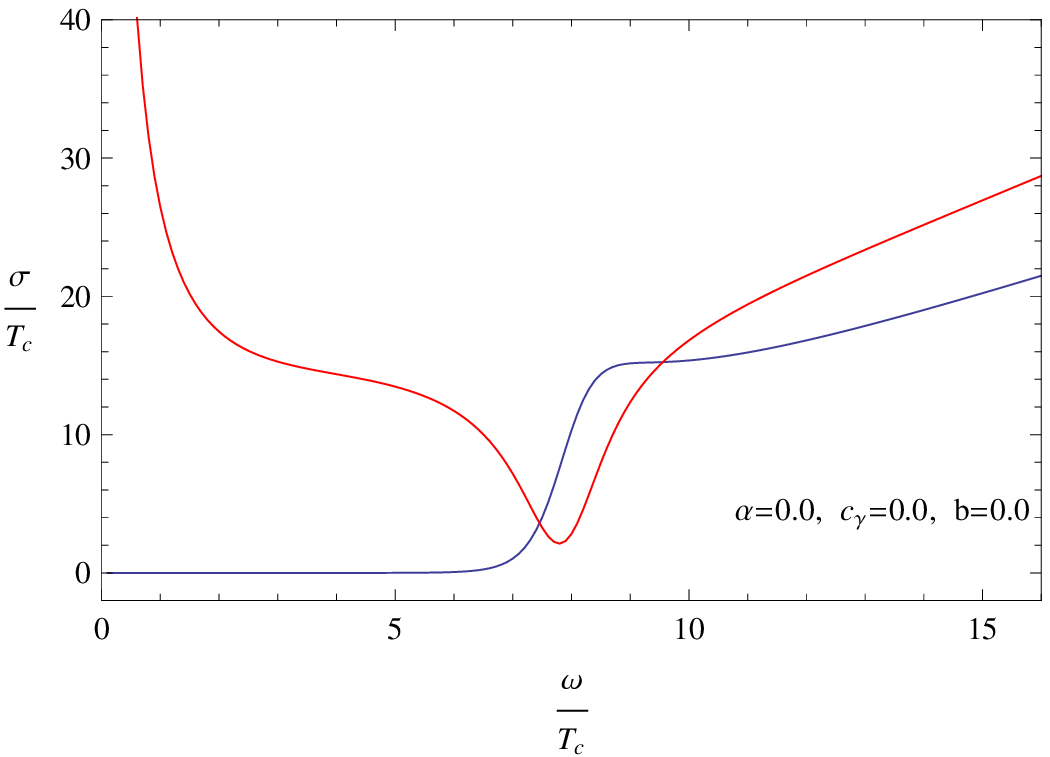}\hspace{0.2cm}%
\includegraphics[scale=0.5]{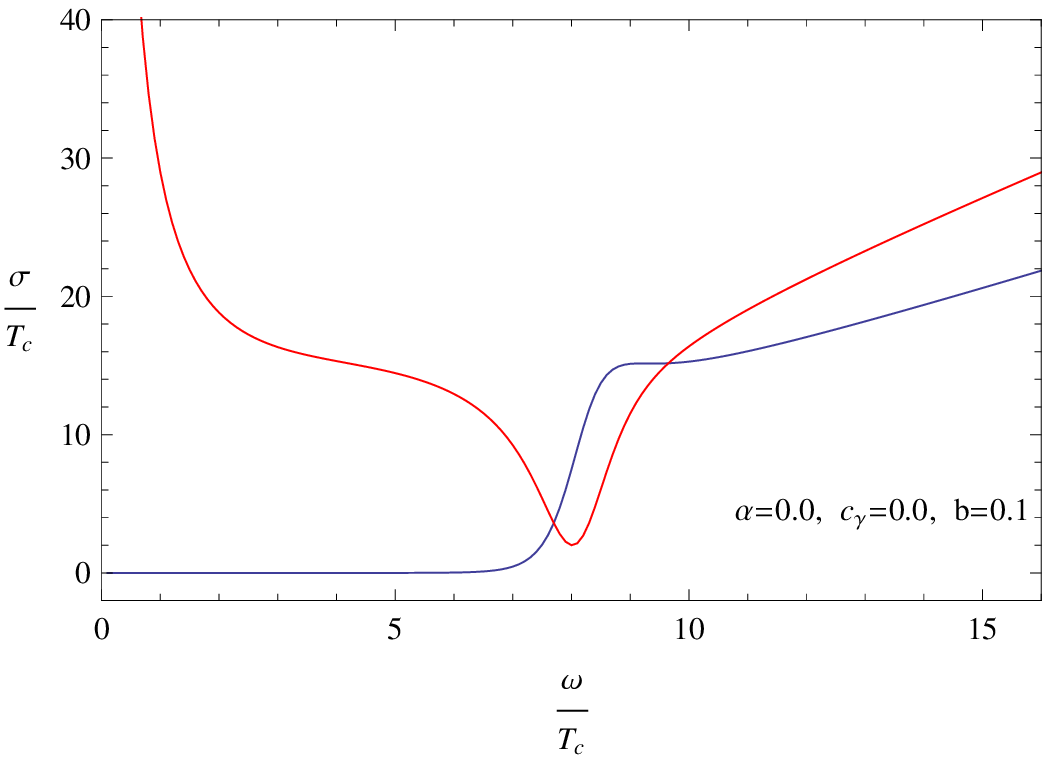}\hspace{0.2cm}%
\includegraphics[scale=0.5]{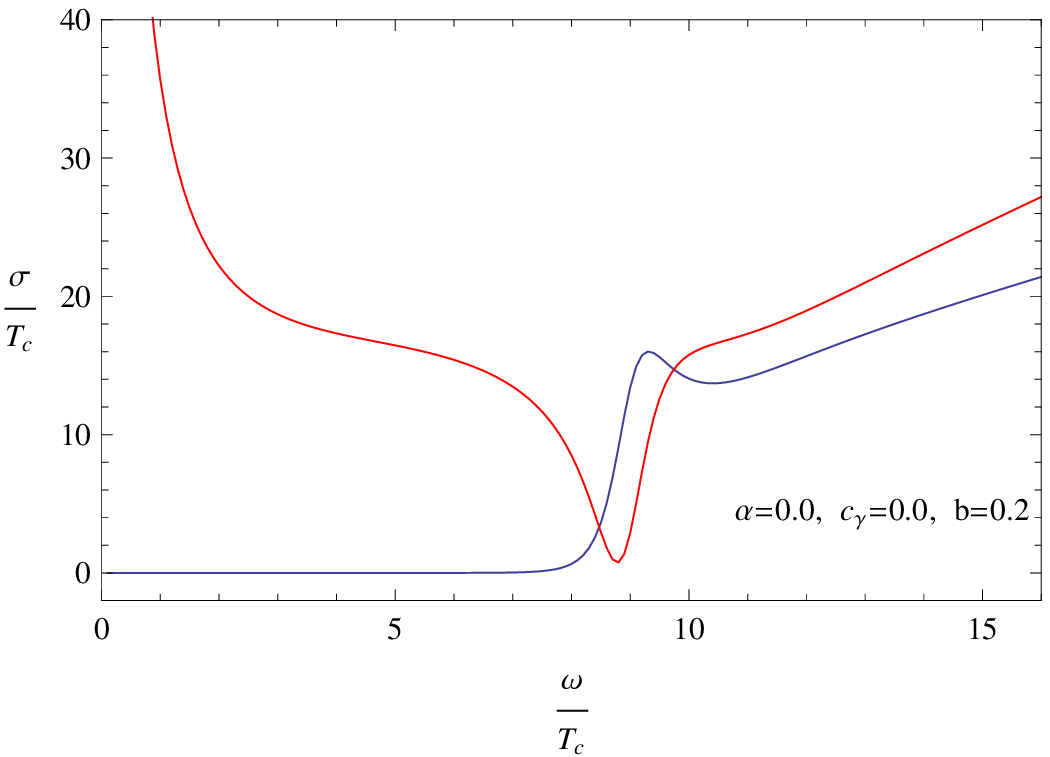}\\ \vspace{0.0cm}
\includegraphics[scale=0.5]{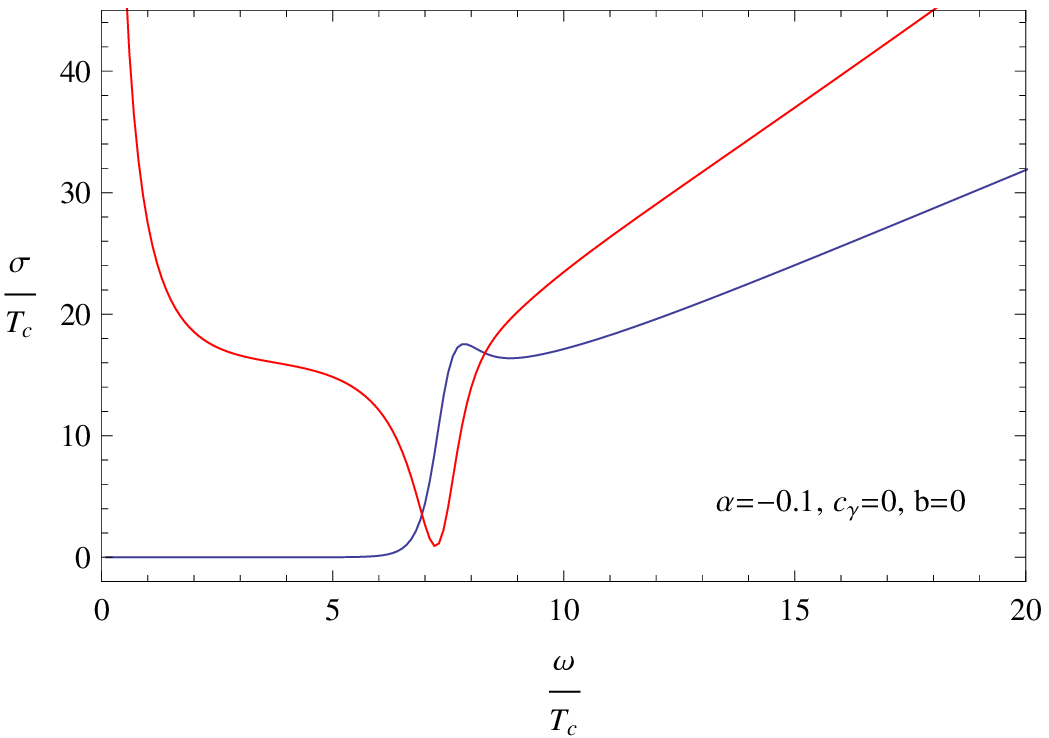}\hspace{0.2cm}%
\includegraphics[scale=0.5]{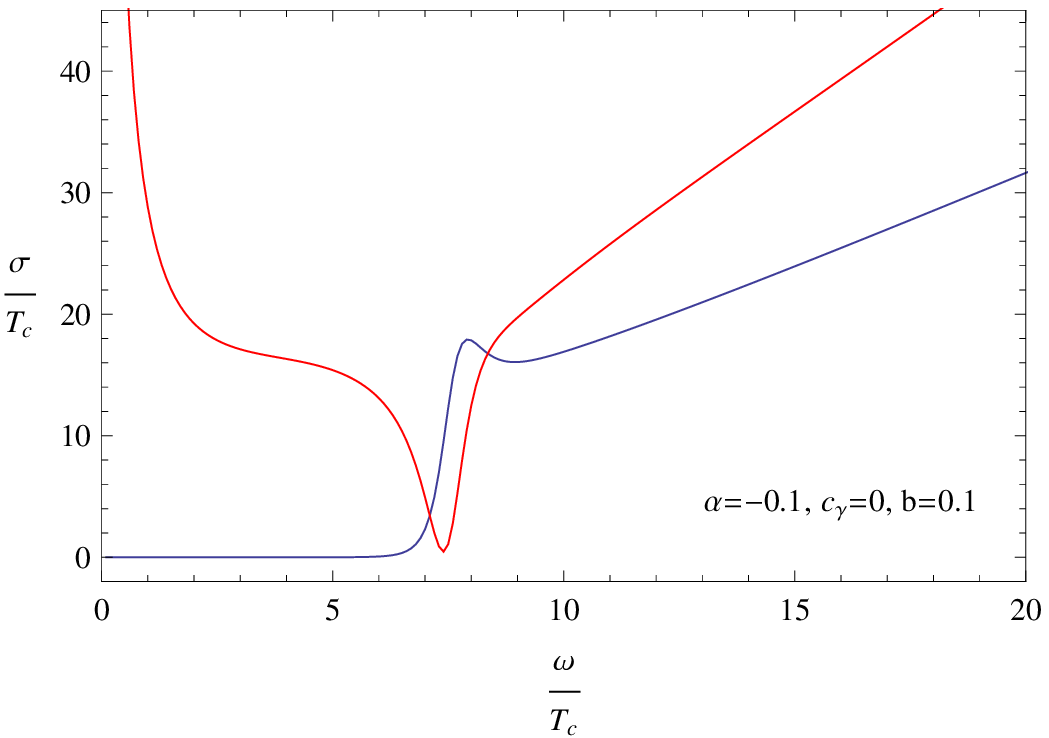}\hspace{0.2cm}%
\includegraphics[scale=0.5]{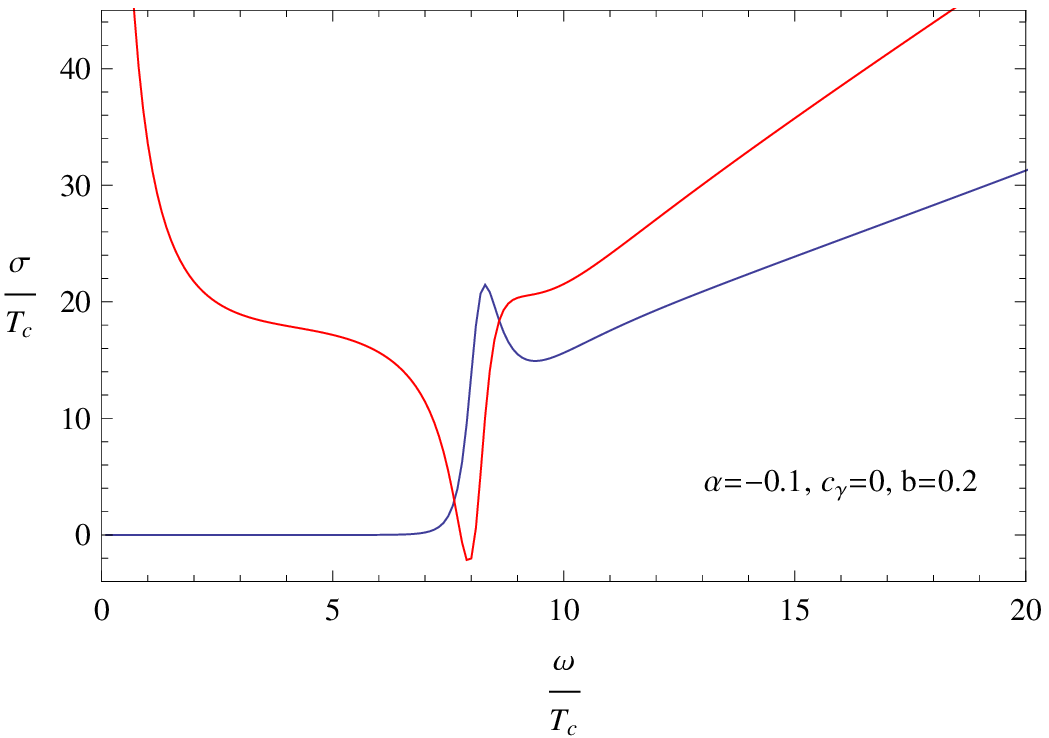}\\ \vspace{0.0cm}

\caption{\label{Conductivity} (Color online) The conductivity of the superconductors as a function of $\omega/T_c$ for different values of $\alpha$ and $b$ with $c_4=0$, $c_\gamma=0$ and $m^{2}L^{2}=-3$. The blue (bottom) line represents the real part of the conductivity, $Re(\sigma)$, and red (top) line is the imaginary part of the conductivity, $Im(\sigma)$.}
\end{figure}

In Fig. \ref{Conductivity} and table \ref{ConductivityTc} we present  the frequency dependent
conductivity obtained by solving the motion equation of the Born-Infeld electrodynamics  numerically
for different values of $\alpha$,  $b$ and $c_\gamma$ with $c_4=0$, $\gamma=3$ and  $m^{2}L_{AdS}^{2}=-3$  (we plot the conductivity at temperature  $T/T_c\simeq  0.3$).  We find that the gap
frequency $\omega_{g}$ increases with the increase of the coupling parameter $b$ for fixed  $\alpha$ and $c_\gamma$,  it decreases as $\alpha$ decreases for fixed $b$ and $c_\gamma$, and it increases as $c_\gamma$ increase for fixed $\alpha$ and $b$. From Figs. \ref{Conductivity} and table  \ref{ConductivityTc}, we find that the ratio of the gap frequency in
conductivity $\omega_g$ to the critical temperature $T_c$ in the Gauss-Bonnet black hole with the Born-Infeld electrodynamics depends on the Gauss-Bonnet constant, the model parameters and the Born-Infeld coupling parameter.

\section{conclusions}

The behaviors of the holographic superconductors in the
Gauss-Bonnet gravity have been investigated
by introducing a complex charged scalar field coupling with an electric field obeyed to Born-Infeld electrodynamics in a planar black-hole background.
We present a detail analysis of the condensation of the operator $\langle{\cal O}_{+}\rangle$ by numerical method. For the interesting simple model $\mathfrak{F}(\psi)=\psi^{2}+c_{4}\psi^{4}$, we know that there is a phase transition from the second order to the first one as we alter the values of the Gauss-Bonnet constant $\alpha$, the model parameter $c_4$ and the Born-Infeld coupling parameter $b$. For the transition point, the relation of $\alpha$, $c_4$ and the critical values $b_c$ and $T_c$ which can separate the first- and second-order behavior is: both $b_c$ and $T_c$ decrease as $\alpha$ increases for fixed $c_{4}$, and $b_c$ decreases but $T_c$ increases as $c_{4}$ increases for fixed $\alpha$. It is interesting to find that the Born-Infeld coupling parameter and model parameters have obvious different effects on the critical temperature for general model  $\mathfrak{F}(\psi)=\psi^{2} +c_{\gamma}\psi^{\gamma}+c_4\psi^4$. If we fix the model parameters ($c_\gamma,~\gamma,~c_4$), we note that the critical temperature becomes smaller as the Born-Infeld coupling parameter $b$ increases for two types of phase transitions, i.e., the scalar hair can be formed harder for the larger $b$. However, the story is completely different if we fix the Born-Infeld coupling parameter $b$. For the cases of second phase transition, the formation of the scalar hair does not affect by model parameters  ($c_\gamma,~\gamma,~c_4$). But for the cases of first phase transition, the scalar hair can be formed easier for the larger model parameter ($c_\gamma,~c_4$) or smaller $\gamma$.
We finally find that the ratio of the gap frequency in
conductivity $\omega_g$ to the critical temperature $T_c$ in the Gauss-Bonnet black hole with the Born-Infeld electrodynamics depends on the Gauss-Bonnet constant $\alpha$, model parameters $(c_4,~c_\gamma, ~\gamma)$, and the coupling parameter $b$. Thus, the Gauss-Bonnet constant, model parameters and Born-Infeld coupling parameter  provide richer physics in  the phase transition and the condensation of the scalar hair.

\begin{acknowledgments}
This work was supported by the National Natural Science Foundation
of China under Grant Nos 10875040, 10905020 and 10875041; a key project of the National
Natural Science Foundation of China under Grant No 10935013; the
National Basic Research of China under Grant No. 2010CB833004,  PCSIRT under Grant No IRT0964, and the Construct Program
of the National Key Discipline. Thanks the Kavali Institute for Theoretical Physics China for hospitality in the final stages of this work.

%S. B. Chen's work was partially
%supported by the National Natural Science Foundation of China under
%Grant No 10875041 and the construct program of key disciplines in
%Hunan Province.

\end{acknowledgments}


\begin{thebibliography}{99}


\bibitem{Maldacena}
J. Maldacena,
%The Large N Limit of Superconformal Field Theories and Supergravity,
 Adv. Theor. Math. Phys. {\bf 2}, 231 (1998).

\bibitem{polyakov} S. S. Gubser, I. R. Klebanov and A. M. Polyakov,
%Gauge Theory Correlators from Noncritical String Theory,
Phys. Lett.
{\bf B 428}, 105 (1998). [hep-th/9802109]

\bibitem{Witten}
E. Witten,
%Anti De Sitter Space And Holography,
Adv. Theor. Math.
Phys. {\bf 2}, 253 (1998).


%\cite{Gubser:2005ih}
\bibitem{Gubser:2005ih}
  S.~S.~Gubser,
%  Phase transitions near black hole horizons,
  Class.\ Quant.\ Grav.\  {\bf 22}, 5121 (2005).
  %[arXiv:hep-th/0505189].
  %%CITATION = CQGRD,22,5121;%%


  \bibitem{GubserPRD78}
S. S. Gubser,
%Breaking an Abelian gauge symmetry near a black hole horizon,
Phys. Rev. D {\bf 78}, 065034 (2008).

\bibitem{HartnollPRL101}
S. A. Hartnoll, C. P. Herzog, and G. T. Horowitz,
%Building an AdS/CFT superconductor,
Phys. Rev. Lett. {\bf 101}, 031601 (2008).


\bibitem{HartnollJHEP12}
S. A. Hartnoll, C. P. Herzog, and G. T. Horowitz,
%Families of IIB duals for nonrelativistic CFTs,
 J. High Energy
Phys. {\bf 0812}, 015 (2008).



\bibitem{HorowitzPRD78}
G. T. Horowitz and M. M. Roberts,
%Holographic Superconductors with Various Condensates,
Phys. Rev. D {\bf 78}, 126008 (2008).

\bibitem{Nakano-Wen}
E. Nakano and Wen-Yu Wen,
%Critical magnetic field in AdS/CFT superconductor,
Phys. Rev. D {\bf 78}, 046004 (2008).

\bibitem{Amado}
I. Amado, M. Kaminski, and K. Landsteiner,
%Hydrodynamics of Holographic Superconductors,
J. High Energy Phys. {\bf 0905}, 021
(2009).

\bibitem{Koutsoumbas}
G. Koutsoumbas, E. Papantonopoulos and G. Siopsis,
%Exact Gravity Dual of a Gapless Superconductor,
J. High Energy Phys. {\bf 0907},
026 (2009).

\bibitem{Maeda79}
K. Maeda, M. Natsuume, and T. Okamura,
%Universality class of holographic superconductors,
Phys. Rev. D {\bf 79}, 126004 (2009).

\bibitem{Sonner}
Julian Sonner,
%A Rotating Holographic Superconductor,
Phys. Rev. D
{\bf 80}, 084031 (2009).

\bibitem{HartnollRev}
S. A. Hartnoll,
%Lectures on holographic methods for condensed matter physics,
Class. Quant. Grav.{\bf 26}, 224002 (2009) [arXiv: 0903.3246.]

\bibitem{HerzogRev}
C. P. Herzog,
%Lectures on Holographic Superfluidity and Superconductivity,
J. Phys. A {\bf 42}, 343001 (2009).

\bibitem{Ammon:2008fc}
  M.~Ammon, J.~Erdmenger, M.~Kaminski, and P.~Kerner,
 % Superconductivity from gauge/gravity duality with flavor,
  Phys.\ Lett.\  B {\bf 680}, 516 (2009).
  %[arXiv:0810.2316 [hep-th]].
  %%CITATION = PHLTA,B680,516;%%

%\cite{Gubser:2009qm}
\bibitem{Gubser:2009qm}
  S.~S.~Gubser, C.~P.~Herzog, S.~S.~Pufu, and T.~Tesileanu,
 % Superconductors from Superstrings,
  Phys.\ Rev.\ Lett.\  {\bf 103}, 141601 (2009).
  %[arXiv:0907.3510 [hep-th]].
  %%CITATION = PRLTA,103,141601;%%


\bibitem{CJ0}
Songbai Chen, Liancheng Wang, Chikun Ding, and Jiliang Jing,
%Holographic superconductors in the AdS black hole spacetime with a
%global monopole,
Nucl. Phys. B{\bf 836}, 222 (2010). [arXiv: 0912.2397]

\bibitem{Gregory}
R. Gregory, S. Kanno, and J. Soda,
%Holographic superconductors with
%higher curvature corrections,
J. High Energy Phys. {\bf 0910}, 010
(2009).


\bibitem{Pan-Wang}
Q. Y. Pan, B. Wang, E. Papantonopoulos, J. Oliveria, and A.B. Pavan, Phys. Rev. D {\bf 81}, 106007 (2010).


\bibitem{Ge-Wang}
X. H. Ge, B. Wang, S. F. Wu, and G. H. Yang, arXiv:1002.4901 [hep-th].

\bibitem{Brihaye}
Y. Brihaye and B. Hartmann, Phys. Rev. D {\bf 81}, 126008 (2010);
arXiv:1003.5130 [hep-th].


\bibitem{Gregory1009}
L. Barclay, R. Gregory, S. Kanno, and P. Sutcliffe,
%Gauss-Bonnet Holographic Superconductors
arXiv:1009.1991[hep-th].

\bibitem{Pan-Wang1} Q. Y. Pan, B. Wang,
%General holographic superconductor models with Gauss-Bonnet corrections
Phys. Lett. B  {\bf 693}, 159   (2010).

\bibitem{Cai-pGB}
Rong-Gen Cai, Zhang-Yu Nie, and Hai-Qing Zhang,
%Holographic p-wave superconductors from Gauss-Bonnet gravity,
Phys. Rev. D{\bf 82}, 066007 (2010); arXiv:1007.3321.
%\cite{Ammon:2008fc}

\bibitem{euler} W. Heisenberg and H. Euler, Z. Phys. 98 (1936)

\bibitem{born} M. Born and L. Infeld, Proc. R. Soc. London, Ser A 144,
425-451 (1934)


\bibitem{gibb} G. W. Gibbons and D. A. Rasheed,
%\emph{Electric-magnetic
%duality rotations in nonlinear electrodynamics},
Nucl. Phys. \textbf{B454},
185 (1995).


\bibitem{Hoffman-Gibbons-Rasheed} B. Hoffmann,
%\emph{Gravitational and electromagnetic mass in the Born-Infeld electrodynamics},
Phys. Rev. \textbf{%
47}, 877 (1935).

\bibitem{Oliveira} H. P. de Oliveira,
%\emph{Nonlinear charged black holes},
Class. Quant. Grav. \textbf{11}, 1469 (1994).

\bibitem{Olivera1} Olivera Mi\v{s}kovi\'{c} and Rodrigo Olea,
%\emph{Conserved charges for black holes in Einstein-Gauss-Bonnet
%gravity coupled to nonlinear electrodynamics in AdS space}
arXiv:1009.5763.

\bibitem{Jing-Chen} Jiliang Jing and Songbai Chen,
%\emph{Holographic superconductors in the Born¨CInfeld electrodynamics}
Phys. Lett. B {\bf 686}, 68 (2010)

\bibitem{Boulware-Deser}
D. G. Boulware and S. Deser, Phys. Rev. Lett. {\bf 55}, 2656
(1985).

\bibitem{Cai-2002}
Rong-Gen Cai, Phys. Rev. D {\bf 65}, 084014 (2002).

%\cite{Charmousis:2002rc}
\bibitem{Charmousis:2002rc}
  C.~Charmousis and J.~F.~Dufaux,
  %``General Gauss-Bonnet brane cosmology,''
  Class.\ Quant.\ Grav.\  {\bf 19}, 4671 (2002).
  %[arXiv:hep-th/0202107].
  %%CITATION = CQGRD,19,4671;%%

\bibitem{FrancoPRD}
S. Franco, A. M. Garcia-Garcia, and D. Rodriguez-Gomez,
%Holographic approach to phase transitions,
Phys. Rev. D {\bf 81}, 041901(R) (2010).
  %[arXiv:0911.1354[hep-th]].
  %%CITATION = PHRVA,D67,104019;%%

\bibitem{Franco}
  S. Franco, A. M. Garcia-Garcia, and D. Rodriguez-Gomez,
%A general class of holographic superconductors,
  JHEP {\bf 04}, 092 (2010).
  %[arXiv:0906.1214[hep-th]].
  %%CITATION = PHRVA,D67,104019;%%



\end{thebibliography}
\end{document}